\documentclass[titlepage]{article}%
\usepackage{scalefnt}%
\usepackage{amsmath}%
\usepackage{amsfonts}%
\usepackage{amssymb}%
\usepackage{graphicx}
\begin{document}

\title{Architecture of optimal transport networks}
\author{Marc Durand\\Mati\`{e}re et Syst\`{e}mes Complexes\\UMR 7057 CNRS \& Universit\'{e} Paris 7 - Denis Diderot\\Tour 33/34 - 2\`{e}me \'{e}tage - case 7056\\2 Place Jussieu - 75251 Paris Cedex 05, France }
\maketitle

\begin{abstract}
We analyze the structure of networks minimizing the global resistance to flow
(or dissipated energy) with respect to two different constraints: fixed total
channel volume and fixed total channel surface area. First, we determine the
shape of channels in such optimal networks and show that they must be straight
with uniform cross-sectional areas. Then, we establish a relation between the
cross-sectional areas of adjoining channels at each junction. Indeed, this
relation is a generalization of Murray's law, originally established in the
context of local optimization. Moreover, we establish a relation between
angles and cross-sectional areas of adjoining channels at each junction, which
can be represented as a vectorial force balance equation, where the force
weight depends on the channel cross-sectional area. A scaling law between the
minimal resistance value and the total volume or surface area value is also
derived from the analysis. Furthermore, we show that no more than three or
four channels meet in one junction of optimal bi-dimensional networks,
depending on the flow profile (e.g.: Poiseuille-like or plug-like) and the
considered constraint (fixed volume or surface area). In particular, we show
that sources are directly connected to wells, without intermediate junctions,
for minimal resistance networks preserving the total channel volume in case of
plug flow regime. Finally, all these results are illustrated with a simple
example, and compared with the structure of natural networks.

\end{abstract}
\date{}

Networked structures arise in a wide array of different contexts such as
water, gas and power supply of a city, vascular systems of plants and animals,
or river basins \cite{Bejan1}\cite{Bejan2}\cite{Changizi}. Thus, optimization
of transport in networks has evident industrial and economical importance, but
may also shed light on the structure of natural networked structures. Indeed,
the analysis of these structures from optimization and selection principles
has been recently the subject of intense scientific activity \cite{West1}%
\cite{West2}\cite{Banavar3}\cite{Banavar4}\cite{Banavar} and controversy
\cite{Kozlowski}\cite{Whitfield}\cite{Banavar2}. Besides, theoretical models -
based on local optimization (i.e. optimization of the geometry of a single
junction) - have been attempted to explain in detail the regular patterns of
vascular networks \cite{Murray}\cite{Zamir}\cite{Zamir2}\cite{Zhi}. However,
it is generally known that as the global optimum is achieved the local optimum
of a single junction is often discarded. In the present paper, we characterize
the structure of networks satisfying to the global optimization of transport.
For the class of networks mentioned here, euclidean metric must be taken
account, and the optimization must be achieved with respect to some
geometrical constraint.

Precisely, the problem we consider can be expressed as it follows: consider
$s$\ sources at the same potential (electrical potential, pressure,
concentration, temperature,...) $V_{S}$\ and $w$\ wells at the same potential
$V_{W}$, their respective positions being fixed. What is the architecture of
the network linking all the sources to all the wells and minimizing the
effective resistance (or dissipated energy), for a fixed total channel volume
or fixed total channel surface area \cite{note1} ? Or equivalently, which
architecture minimizes the total channel volume or surface area for a same
value of the global resistance ?

In the following, we shall refer often to the electrical circuit terminology,
although this study obviously concerns any flow-in-network situation. Let us
denote each pipe by a pair of indices $(i,j)$\ corresponding to the labels of
its two ends. We suppose \textit{a priori} that pipes can be curved, but we
assume that their aspect ratios are sufficiently high so a length $l_{ij}%
$\ and a local cross-sectional area $s_{ij}(l)$\ (where $l$\ denotes the
curvilinear coordinate along a channel) can be unequivocally defined for each
pipe $(i,j)$. The resistance $dr_{ij}$\ of an infinitesimal piece of pipe of
length $dl$\ is then defined as:
\begin{equation}
dr_{ij}=\frac{\rho}{s_{ij}^{m}}dl,
\end{equation}
where $\rho$\ is the "resistivity", supposed to be the same for all the pipes.
For $m=1$, the flow in each channel is plug-like, while for $m=2$\ the flow is
Poiseuille-like. Assuming there is no leakage through the pipe lateral
surface, the resistance of the whole pipe $(i,j)$\ is:%
\begin{equation}
r_{ij}=%
{\displaystyle\int\limits_{0}^{l_{ij}}}
\frac{\rho}{s_{ij}^{m}}dl.
\end{equation}
Since we shall inspect the minimal resistance configuration with respect with
two different constraints (a fixed total channel volume $V_{tot}$\ and a fixed
total surface-area channel $S_{tot}$), we introduce for simplicity the
"constraint function": $C_{n}=%
{\displaystyle\sum\limits_{\left(  i,j\right)  }}
{\displaystyle\int\limits_{0}^{l_{ij}}}
s_{ij}^{n}dl$\ , so that: $C_{1}=V_{tot}$, and $C_{1/2}\propto S_{tot}$.

\section{Cohn's theorem}

To characterize the architecture of minimal resistance networks, we shall
invoke Cohn's theorem, originally developed in the context of electrical
circuit analysis \cite{Cohn}: consider a one-port network composed entirely of
two-terminal elements with resistances $r_{ij}$. The variation of the
effective network resistance $R$\ with the variation of the resistance
$r_{ij}$\ is given by:%
\begin{equation}
\frac{\partial R}{\partial r_{ij}}=\left(  \frac{i_{ij}}{I}\right)
^{2}.\label{Cohn}%
\end{equation}
No particular assumption is made on the expression of the resistances $r_{ij}
$\ for the derivation of this result (indeed, the theorem is still valid for
complex impedances). Conservation of flow and energy only are required. Thus,
Cohn's theorem can be applied to a broader class of flow-in-network situations.

\section{Optimal shape of channels}

We first notice that in order for the effective network resistance to be at
its minimum value with respect to the constraint $C_{n}$, each channel must be
straight with a uniform cross-sectional area (i.e.: $s_{ij}(l)=s_{ij}$).
Indeed, we see from eq. \ref{Cohn} that the effective network resistance
$R$\ is a monotone function of the individual resistances $r_{ij}$. Thus, any
small change in pipe diameter or pipe length from the minimal resistance
configuration - compatible with the constraint - must lead to an increase of
the resistances $r_{ij}$. As a consequence, the length of each pipe must be as
small as possible and its diameter as large as possible, i.e. each channel
must be straight with a uniform cross-sectional area. Besides, it can be
noticed that a circular cross-sectional area have the specific property of
minimizing both the pipe surface area for a fixed volume (or equivalently
maximizing the pipe volume for a fixed surface area) and the dissipative
energy in the channel for a fixed incoming flow-rate in case of
Poiseuille-flow regime.

\section{Relations between diameters: generalized Murray's law}

\begin{figure}
[h]
\begin{center}
\includegraphics[
natheight=2.633400in,
natwidth=3.665400in,
height=1.8198in,
width=2.5264in
]%
{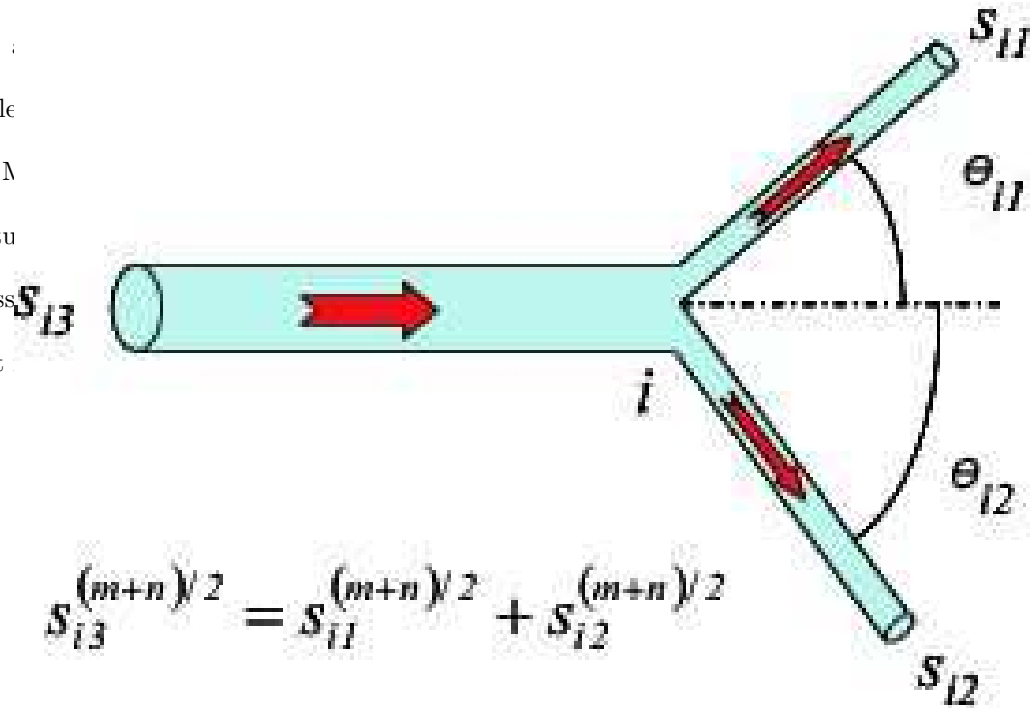}%
\caption{Relation between cross-sectional areas of adjoining channels in a
minimal resistance network. This relation is a generalization of Murray's
law.}%
\label{equilibrium1}%
\end{center}
\end{figure}

We now establish relations between diameters and angles in an optimal network,
for a given topology (meaning that no junction or channel can be added or
removed from the network, but the channel lengths and cross-section areas are
free to vary). We thus have to minimize the function $\widetilde{R}=R+\lambda
C_{n}$\ (where $\lambda$\ is a Lagrange multiplier) with respect to the
independent variables $\left\{  s_{ij}\right\}  $\ and $\left\{
\mathbf{r}_{i}=\left(  x_{i},y_{i}\right)  \right\}  $, respectively the
channel cross-sectional areas and node positions. Using Cohn's theorem
\ref{Cohn}, the condition of extremum with respect to the cross-sectional
areas ($\partial\widetilde{R}/\partial s_{ij}=0$) gives:%

\begin{equation}
\left(  \frac{i_{ij}}{I}\right)  ^{2}=\frac{\lambda}{\rho}\frac{n}{m}%
s_{ij}^{m+n}.\label{currents}%
\end{equation}
Furthermore, conservation of flow-rate at each junction $i$ ($%
{\displaystyle\sum\limits_{j}}
i_{ij}=0$) implies:%
\begin{equation}%
{\displaystyle\sum\limits_{j}}
sign\left(  i_{ij}\right)  s_{ij}^{\left(  m+n\right)  /2}=0.\label{diameters}%
\end{equation}
This relation, illustrated on Fig. \ref{equilibrium1} and valid for
netted-like as for tree-like networks, is a generalization of Murray's law
\cite{Murray} to any flow profile and with different constraints (Murray's law
was originally derived for the particular case $m=2$, $n=1$). Moreover, we
must point out that relation \ref{diameters} results here from the global
optimization of the network structure, while the original derivation of
Murray's law was based on a local optimization (flow and channel
cross-sectional area were functionally related: an optimal cross-sectional
area was found for a given flow, and not for all levels of total flow).%

\section{Geometry of nodes}

\begin{figure}
[h]
\begin{center}
\includegraphics[
natheight=2.633400in,
natwidth=3.663700in,
height=1.8497in,
width=2.567in
]%
{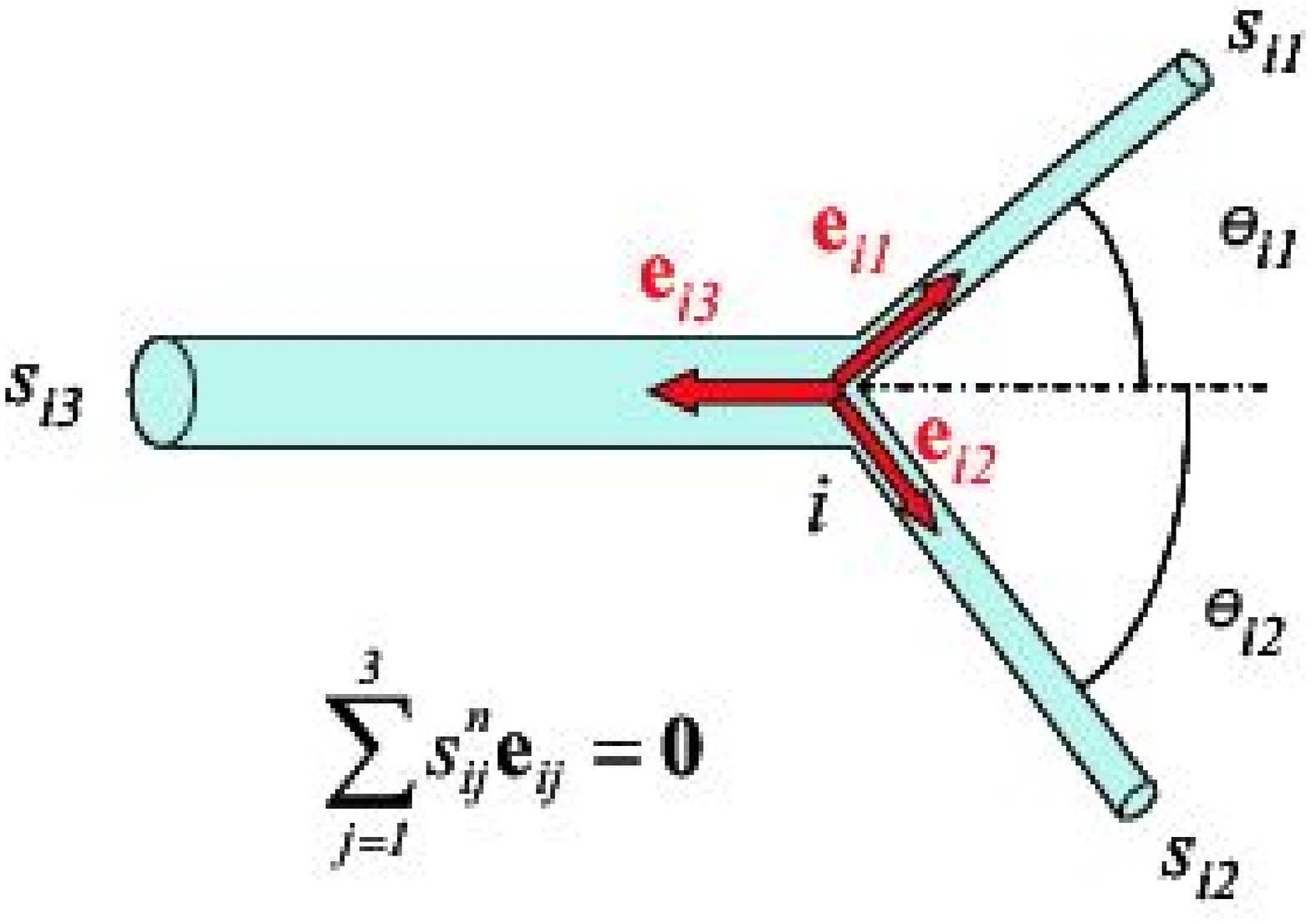}%
\caption{Relation between angles and cross-sectional areas of adjoining
channels in a minimal resistance network. This relation is similar to a force
balance equation describing the equilibrium of strings tied together and under
respective tensions, or weights, $s_{ij}^{n}$.}%
\label{equilibrium2}%
\end{center}
\end{figure}

Condition of extremum with respect to the node positions ($\partial
\widetilde{R}/\partial r_{i}=0$) together with relation \ref{currents}
straightforwardly leads to the following vectorial equality at each node $i$:
\begin{equation}%
{\displaystyle\sum\limits_{j}}
s_{ij}^{n}\mathbf{e}_{ij}=\mathbf{0,}\label{angles}%
\end{equation}
where $\mathbf{e}_{ij}$\ is the outward-pointing unit vector along the channel
$\left(  i,j\right)  $\ (see Fig. \ref{equilibrium2}). This equality, relating
angles between adjoining channels to their cross-sectional areas, is similar
to a force balance equation, where the weight of the force acting along the
channel $\left(  i,j\right)  $\ is directly proportional to $s_{ij}^{n}$. As
for Murray's law, local optimization principles have already been proposed in
order to describe the geometry of nodes in natural networks, namely:
minimization of channel volume (V), channel surface area (S), dissipated power
(P), and drag force (D) on the walls \cite{Changizi}\cite{Zamir}\cite{Zamir2}.
All these approaches consist in varying the position of a given junction,
while the positions of the other junctions, the network topology, the channel
cross-sectional areas, and the flow-rates through every channel are remained
fixed. However, in the context of a global optimization, a change in a node
position should alter the flow-rate distribution, and it is therefore to be
expected that global minimization of the dissipated energy leads to a
different optimal geometry of nodes than in the local optimization context
(P). Indeed, the optimal geometry of nodes described by Eq. \ref{angles} is
similar to the one obtained for (S) (when $n=1/2$) or (V) (when $n=1$), but
different from (P) \cite{Changizi}\cite{Zamir}\cite{Zamir2}.%

\section{Scaling-law between minimal resistance and constraint value}

A relation between the minimal resistance value and the constraint value can
be established, using Eq. \ref{currents} and conservation of energy:%
\begin{equation}
R=%
{\displaystyle\sum\limits_{\left(  i,j\right)  }}
r_{ij}\left(  \frac{i_{ij}}{I}\right)  ^{2}=\lambda\frac{n}{m}C_{n}%
.\label{R-Cn}%
\end{equation}
On the other hand, a classical result of optimization theory relates the
Lagrange multiplier to the change of the minimal resistance with respect to
the constraint value: $\lambda=-\frac{dR_{m}}{dCn}$\ (note that Eq. \ref{R-Cn}
implies $\lambda\geq0$). Therefore, it is found that the resistance of an
optimal network scales as $C_{n}^{-m/n}$, i.e.:
\begin{equation}
R=\rho l\left(  \frac{l}{C_{n}}\right)  ^{\frac{m}{n}},\label{scaling}%
\end{equation}
where $l$\ is a parameter with dimension of length, depending solely on the
network topology, the positions of sources and wells, and the values of $m$
and $n$.

We have shown that a minimal resistance configuration, for a given topology,
if it does exist, must satisfy to the equations \ref{currents},
\ref{diameters}, \ref{angles}, and \ref{scaling}. Wether the extrema
characterized by this set of equations are local minima or local maxima is not
clear (although this uncertainty might be dispelled by some convexity
argument). Nevertheless, because individual resistances have finite values,
there must exist at least one configuration with global minimal resistance
(but we do not know if this configuration is unique) \cite{note2}.

\section{Upper bound on the node connectivity}

\begin{figure}
[h]
\begin{center}
\includegraphics[
natheight=2.711500in,
natwidth=7.369800in,
height=1.3267in,
width=3.5816in
]%
{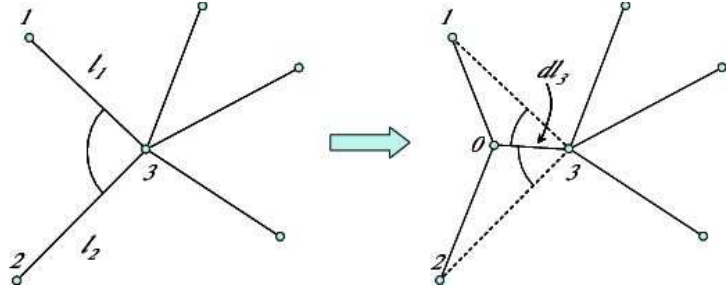}%
\caption{Elementary transformation of a $N$-fold junction to a $\left(
N-1\right)  $-fold junction plus a $3$-fold junction. A new channel, with
infinitesimal length $dl_{3}$ is thus created.}%
\label{decomposition}%
\end{center}
\end{figure}

Finally, we establish an upper bound on the number of channels joining in one
node, in a bi-dimensional minimal resistance network. To do so, we look at a
given junction of $N$\ channels and determine when this junction is
preferentially replaced with two junctions respectively of $3$\ and
$N-1$\ channels. Suppose we create a new channel of infinitesimal length
$dl_{3}$, as depicted in Fig. \ref{decomposition}. Then, the length variation
of the two other channels joining in the new $3$-fold junction are:
$dl_{1}=-dl_{3}\cos\theta_{1}$\ and $dl_{2}=-dl_{3}\cos\theta_{2}$, with:
$\theta_{1}+\theta_{2}=\gamma$, where $\gamma$\ is the angle between these two
adjacent channels. The variation of the associated resistances are
respectively: $dr_{1}=-\rho dl_{3}\cos\theta_{1}/s_{1}^{m}$, $dr_{2}=-\rho
dl_{3}\cos\theta_{2}/s_{2}^{m}$, and $dr_{3}=\rho dl_{3}/s_{3}^{m}$, where
$s_{1}$, $s_{2}$\ and $s_{3}$\ are the respective channel cross-sectional
areas.\ Moreover, this transformation must preserve the value of $C_{n}$, so
the new channel cross-sectional area $s_{3}$\ must satisfy:
\begin{equation}
s_{3}^{n}=s_{1}^{n}\cos\theta_{1}+s_{2}^{n}\cos\theta_{2}.\label{change}%
\end{equation}
Using once again Cohn's theorem, we obtain the variation of the effective
resistance:%
\begin{equation}
dR=\rho\frac{dl_{3}}{I^{2}}\left(  \frac{i_{3}^{2}}{s_{3}^{m}}-\frac{i_{1}%
^{2}\cos\theta_{1}}{s_{1}^{m}}-\frac{i_{2}^{2}\cos\theta_{2}}{s_{2}^{m}%
}\right)  .\label{dR}%
\end{equation}%
Suppose now that the $N$-fold junction was in a minimal resistance
configuration. Then, conditions \ref{currents} and \ref{angles} must be
fulfilled, and we can replace $i_{1}^{2}$\ and $i_{2}^{2}$\ in Eq. \ref{dR} by
their expressions (Eq. \ref{currents}). Moreover, conservation of flow rate
relates $i_{3}$\ to $i_{1}$\ and $i_{2}$: $i_{3}=-i_{1}-i_{2}$. Using Eq.
\ref{change}, we see that the resistance variation $dR$\ is negative when:
$s_{3}^{\left(  m+n\right)  /2}\geq s_{1}^{\left(  m+n\right)  /2}\pm
s_{2}^{\left(  m+n\right)  /2}$. The sign in the right-hand side of this
inequality is positive when the two adjacent channels are crossed by flows in
same direction, and negative when they are crossed by flows in opposite
directions. The former inequality can be rewritten as: $\cos\theta_{1}%
+r^{n}\cos\theta_{2}\geq\left(  1\pm r^{\frac{m+n}{2}}\right)  ^{\frac
{2n}{m+n}}$, with $r=s_{2}/s_{1}$. Before establishing an upper-bound on the
node connectivity, we must notice the following "rules" on the geometry of
junctions in optimal networks:

\begin{enumerate}
\item There is at least one angle lower than $2\pi/N$\ between two adjacent
channels in a $N$-fold junction (from geometrical consideration).

\item There is at least one pair of adjacent channels crossed by flows in
opposed directions (from flow conservation).

\item The angle between two adjacent channels is always lower than $\pi
$\ (from Eq. \ref{angles}).
\end{enumerate}

Let us choose $\theta_{1}$\ and $\theta_{2}$\ such that $\sin\theta_{1}%
=r^{n}\sin\theta_{2}$, what corresponds to the maximum value of the left-hand
side of the former equality. Since $\gamma\leq\pi$\ (rule 3.), we easily check
that both $\theta_{1}$\ and $\theta_{2}$\ are positive and lower than $\pi/2$,
and simple algebra leads to:
\begin{equation}
\cos\theta_{1}+r^{n}\cos\theta_{2}=\sqrt{1+r^{2n}+2r^{n}\cos\gamma}.
\end{equation}
Thus, the resistance variation $dR$\ is negative if and only if:%
\begin{equation}
\cos\gamma\geq f_{\pm}\left(  r\right)  =\frac{\left(  1\pm r^{\frac{m+n}{2}%
}\right)  ^{\frac{4n}{m+n}}-1-r^{2n}}{2r^{n}},
\end{equation}
where the functions $f_{+}\left(  r\right)  $\ and $f_{-}\left(  r\right)
$\ correspond to the respective situations of two adjacent channels crossed by
flows in same and opposite directions. The analysis of $f_{+}\left(  r\right)
$\ and $f_{-}\left(  r\right)  $\ shows that, for any value of $r$, these
functions are bounded in the following way: $f_{+}\left(  r\right)
\leq2^{\frac{3n-m}{m+n}}-1$\ for any value of $m$, $n$; $f_{-}\left(
r\right)  \leq0 $\ if $m>n$, and $\ f_{-}\left(  r\right)  \leq-1$\ if $m=n$.
So if $\gamma$\ is lower than $\gamma_{+}=\arccos\left(  2^{\frac{3n-m}{m+n}%
}-1\right)  $\ for the first situation, or $\gamma_{-}=90%
{{}^\circ}%
$\ (if $m>n$) or $180%
{{}^\circ}%
$\ (if $m=n$) for the second situation, we are ensured that the resistance
variation is negative. Let us inspect the different situations:

\begin{itemize}
\item If $m=2$, and $n=1/2$: $\gamma_{+}\simeq97.4%
{{}^\circ}%
$, $\gamma_{-}=90%
{{}^\circ}%
$. We know there is at least one angle lower than $360%
{{}^\circ}%
/N$\ between two adjacent channels in a $N$-fold junction (rule 1.). By
choosing this angle as $\gamma$\ in the previous analysis, we conclude that a
$N$-fold junction is preferably replaced with a $\left(  N-1\right)  $-fold
junction plus a $3$-fold junction, as long as $N\geq4$. The new structure is
not in a minimal resistance configuration, determined by Eqs \ref{currents}
and \ref{angles}, so the "relaxation" of the new structure such a
configuration implies a further decrease of the effective resistance.
Eventually, we can repeat the same reasoning on the $\left(  N-1\right)
$-fold junction, if $N-1\geq4$. We come to the conclusion that exactly three
channels meet at each junction in such an optimal network.

\item If $m=2$, and $n=1$\ or $m=1$, and $n=1/2$: $\gamma_{+}\simeq74.9%
{{}^\circ}%
$, $\gamma_{-}=90%
{{}^\circ}%
$. Following the same argumentation, we conclude that a $N$-fold junction is
preferably replaced with a $\left(  N-1\right)  $-fold junction plus a
$3$-fold junction as long as $N\geq5$. Thus, no more than four channels meet
in one junction in such an optimized network. Furthermore, it can be noticed
that only two kinds of $4$-fold junctions can exist in such a network: either
three adjacent channels are crossed by flows of same sign (and the last flow
has an opposed sign), or two adjacent channels are crossed by flows with same
sign and the two other adjacent channels are crossed by flows with same
opposite sign. A $4$-fold junction with channels crossed by flows with
alternate signs is preferably replaced by two $3$-fold junctions, since there
is always two adjacent channels crossed by flows with opposite signs and with
an angle lower than $90%
{{}^\circ}%
$\ (rule 1.).

\item If $m=1$\ and $n=1$: $\gamma_{+}=0%
{{}^\circ}%
$, $\gamma_{-}=180%
{{}^\circ}%
$. But we know that there is always two adjacent channels crossed by flows
with opposite signs in a $N$-fold junction (rule 2.), with an angle between
them lower than $180%
{{}^\circ}%
$\ (rule 3.). So the $N$-fold junction is preferably replaced with a $\left(
N-1\right)  $-fold junction plus a $3$-fold junction for any $N\geq4$. Now, if
we let the new structure of the network "relax" to a minimal resistance
configuration, it must simultaneously satisfy Eqs. \ref{diameters} and
\ref{angles} at every junction, and particularly at the $3$-fold junction. But
this set of equations applied in a $3$-fold junction has only trivial
solutions when $m\leq n$: either one cross-section is null, or the three
channels are colinear. We conclude that sources are directly connected to the
wells, with no intermediate junction, in a minimal resistance network
preserving total channel volume and in case of plug-flow regime.
\end{itemize}

As a concluding remark for this section, we point out that the same reasoning
may be used on the total channel length variation instead of resistance
variation (Steiner tree problem). In that case, we obtain that links meet at
threefold junctions (with equal angles of $120%
{{}^\circ}%
$) in a length-minimizing network.

\section{A simple example}

\begin{figure}
[h]
\begin{center}
\includegraphics[
natheight=5.209600in,
natwidth=5.968400in,
height=2.8867in,
width=3.3043in
]%
{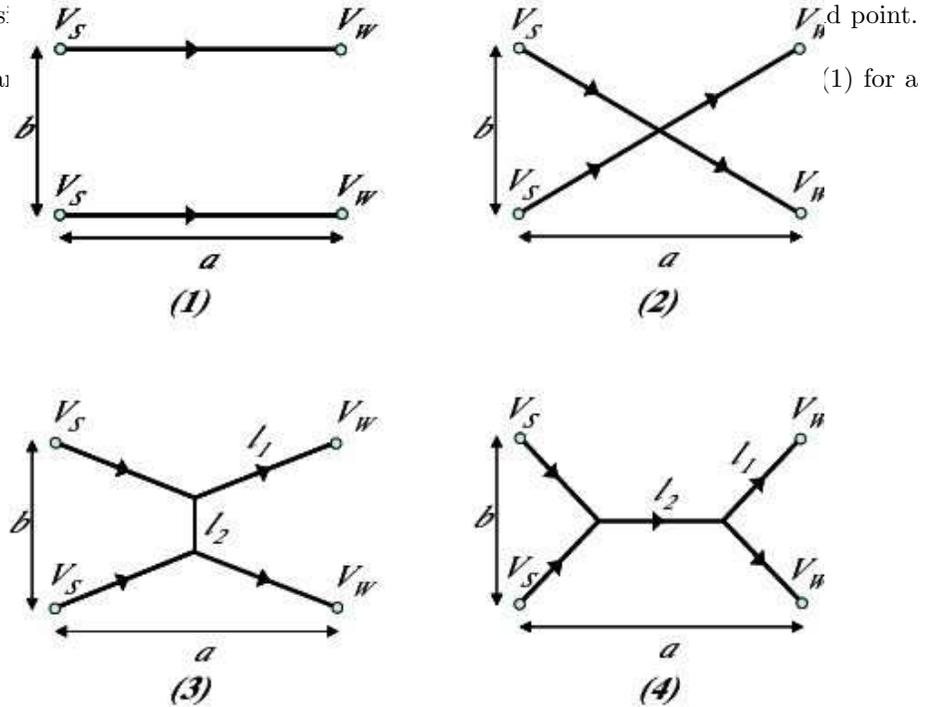}%
\caption{Four different network configurations linking two sources to two
wells, placed at the corners of a rectangle of length $a$ and $b$.}%
\label{example}%
\end{center}
\end{figure}

We compare our results with a simple example: two sources and two wells placed
at the corner of a rectangle, as depicted on Fig. \ref{example}. Four
configurations are analyzed. In configuration (1), sources are directly
connected to wells, without any intermediate junction. In Configuration (2),
sources are connected to wells \textit{via} a $4$-fold intermediate junction.
In configurations (3) and (4), sources and wells are connected through two
$3$-fold junctions (and the position of intermediate junctions are chosen such
that equality \ref{angles} is satisfied). The corresponding dimensionless
resistance $\frac{R}{\rho a}\left(  \frac{C_{n}}{a}\right)  ^{\frac{m+n}{n}}%
$of each configuration is reported on table \ref{Table}. From these
expressions, we note the following observations, in agreement with our
results: firstly, we notice that $R$\ scales as $\left(  1/C_{n}\right)
^{m/n} $. Secondly, when $m=n$\ $\left(  =1\right)  $, configuration (1) is
the smallest resistance configuration, for any value of the aspect ratio $b/a
$. Thirdly, resistance of configuration (3) is always lower than resistance of
configuration (2) and higher than resistance of configuration (1) ($R_{1}\leq
R_{3}\leq R_{2}$), for any value of $m$, $n$, and $b/a$. Fourthly, resistance
of configuration (4) is lower than resistance of configuration (2) as soon as:
$\sqrt{1+\left(  b/a\right)  ^{2}}\leq2^{2(m-n)/(m+n)}$, for any value of $m$,
$n$, and $b/a$. One can easily check that this criterion on the aspect ratio
$b/a$\ (for given values of $m$\ and $n$) corresponds to the condition for Eq.
\ref{diameters} to be simultaneously satisfied with Eq. \ref{angles} at each
$3$-fold junctions of configuration (4). In particular, resistance of
configuration (4) cannot be lower than resistance of configuration (2) when
$m=n$, in agreement with the second point. Fifthly, when $m>n$, resistance of
configuration (4) can be lower than resistance of configuration (1) for a
sufficiently low value of $b/a$.%

\begin{table}[htbp] \centering
$%
\begin{tabular}
[c]{|c|p{5.65cm}|}\hline
\textbf{Configuration} & \multicolumn{1}{|c|}{$\frac{R}{\rho a}\left(
\frac{C_{n}}{a}\right)  ^{\frac{m}{n}}$}\\\hline
(1) & \multicolumn{1}{|c|}{$2^{\frac{m-n}{n}}$}\\\hline
(2) & \multicolumn{1}{|c|}{$2^{\frac{m-n}{n}}\left(  1+\left(  \frac{b}%
{a}\right)  ^{2}\right)  ^{\frac{m+n}{n}}$}\\\hline
(3) & $\widehat{l}_{1}\left(  4\widehat{l}_{1}+2\widehat{l}_{2}\frac
{b/a-\widehat{l}_{2}}{\sqrt{1+\left(  b/a-\widehat{l}_{2}\right)  ^{2}}%
}\right)  ^{\frac{m}{n}}$, \ \ \ \ \ \ \ \ \ with $\widehat{l}_{1}=\frac
{\sqrt{1+\left(  b/a-\widehat{l}_{2}\right)  ^{2}}}{2}$\\\hline
(4) & $\left(  4\widehat{l}_{1}+\widehat{l}_{2}\frac{1-\widehat{l}_{2}%
}{\widehat{l}_{1}}\right)  ^{\frac{m}{n}}\left(  \widehat{l}_{1}+\widehat
{l}_{2}\left(  \frac{\widehat{l}_{1}}{1-\widehat{l}_{2}}\right)  ^{\frac{m}%
{n}}\right)  $, with $\widehat{l}_{1}=\frac{\sqrt{\left(  b/a\right)
^{2}+\left(  1-\widehat{l}_{2}\right)  ^{2}}}{2}$\\\hline
\end{tabular}
$%
\caption{Dimensionless resistances $\frac{R}{\rho a}\left( \frac{C_{n}}{a}\right) ^{\frac{m}{n}}$ corresponding to the four configurations depicted on Fig.  \ref{example}. For configurations (3) and (4), $\widehat{l}_{1}=l_{1}/a$
and $\widehat{l}_{2}=l_{2}/a$ are the dimensionless lengths of the two kind of channels.}\label{Table}%
\end{table}%

\section{Comparison with natural networks}

\begin{figure}
[h]
\begin{center}
\includegraphics[
height=2.1967in,
width=3.0867in
]%
{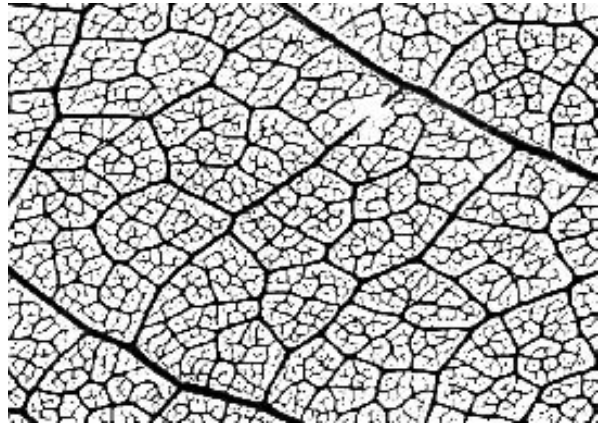}%
\caption{Portion of leaf venation. In most species, the structure is
netted-like, and veins meet in $3$-fold junctions.}%
\label{leaf}%
\end{center}
\end{figure}

All the results derived in previous sections (relations \ref{currents},
\ref{diameters}, \ref{angles}, \ref{scaling} as well as the upper-bound on the
node connectivity) are consequences of global optimization. However, these
results have been established by studying any local perturbation of the
structure. Such a local adaptive process may take place during ontogeny of
natural networks. Therefore, it may be of interest to compare the structure of
some natural networks with the results presented in this work. Indeed, it has
been already shown in various publications \cite{Sherman}\cite{LaBarbera} that
Murray's law is well satisfied in some appropriate portions of human and
animal vascular systems. In that case, the flow profile is nearly
Poiseuille-like ($m=2$) and the relevant constraint is a fixed total channel
volume ($n=1$) \cite{Sherman}\cite{LaBarbera}. Validity of Murray's law for
vascular system of plants is more controversial \cite{LaBarbera}%
\cite{Roth}\cite{Canny}\cite{McCulloh}, mostly because of the underlying
theoretical assumptions in the original derivation of Murray's law, and of the
particular structure of veins in vascular system of plants. Nevertheless,
experimental data suggest that a relation $%
{\displaystyle\sum\limits_{j}}
sign\left(  i_{ij}\right)  s_{ij}^{\nu/2}=0$\ is still verified, with $\nu
$\ between $2.49$\ and $3$ \cite{Zhi}\cite{McCulloh}\cite{Kruszewski}%
\cite{Kizilova}. Let us look more precisely at the bi-dimensional leaf
venation network, like the one reported on Fig. \ref{leaf}. Leaf veins are
actually vascular bundles \cite{Xylems}, supporting two parallel flows : a
pressure-driven flow of water and minerals from petiole to stomata through
xylem tissues, and a diffusive flow of nutrients and photosynthesis products
in the opposite direction through phloem tissues. So petiole (or major vein)
and stomata play alternatively roles of sources and wells for the leaf. An
outer layer of cells, called the bundle sheath, surrounds the vascular
tissues. Although this layer is not fully impermeable, the leaky radial flow
is small when compared with the axial flow, except for the minor veins
\cite{Roth}\cite{Holbrook}. For these veins, leakage is very important and the
pressure field and nutrient concentration nearby are almost uniform. Kull and
Herbig \cite{Kull} investigated on leaf topology of several species. They
observed that leaf venations preferably show trivalent nodes with six
neighbors, and noticed that this geometry is typical of self-generating
structures like bubble floats. In a recent study, S. Bohn \textit{et al.}
\cite{Bohn} analyzed geometry of junctions in the leaf venation of various
species. They observed that angles between veins are very well defined and
that a vectorial balance equation comparable to eq. \ref{angles} can be
established, where the weight of each vector is directly proportional to the
vein radius (i.e. $n=1/2$). Comparison of Bohn \textit{et al.} observations
with our optimization principles suggest then that structure of leaf venation
corresponds to a minimization of the resistance for a fixed total channel
surface area (or minimization of surface area for a fixed value of
resistance). This result is coherent with the idea of a predominant building
cost of the bundle sheath cells over those of the vascular tissues
\cite{Roth}\cite{McCulloh}. Taking $n=1/2$\ and comparing Eq. \ref{diameters}
with experimental studies of Murray's law leads to a value of $m$\ between
$1.99$\ and $2.5$, meaning that flow in veins is nearly Poiseuille-like. Note
it is assumed in the theory that all channels have same resistivity $\rho$.
However, density of xylem and phloem tissues in a leaf vein might be a
function of the vein diameter as well. This variation of the resistivity is
then include in the coefficient $m$, what could explain the slight difference
observed between the experimental value of $m$ and the theoretical value for a
Poiseuille-flow regime. All these results and the presence of trivalent nodes
suggest then that structure of leaf venation correspond to minimal resistance
configuration preserving the total surface area and nearly Poiseuille flow
profile. The measure of the scaling-law between the hydraulic resistance and
the total channel volume or surface area might be an additional way to test
this conjecture.%

\begin{thebibliography}{99}                                                                                               %
\bibitem {Bejan1}A.\ Bejan, \textit{Rev. G\'{e}n. Therm.} \textbf{36}, 592-604 (1997).

\bibitem {Bejan2}S. Lorente, W.\ Wechsatol, and A.\ Bejan, \textit{Int. J.
Heat and Mass Transfer} \textbf{45}, 3299-3312 (2002).

\bibitem {Changizi}M. A. Changizi and C. Cherniak, \textit{Can. J. Physiol.
Pharmacol.} \textbf{78}, 603-611 (2000).

\bibitem {West1}G.\ B.\ West, J.\ H.\ Brown, and B.\ J.\ Enquist,
\textit{Science} \textbf{276}, 122-126 (1997).

\bibitem {West2}G.\ B.\ West, J.\ H.\ Brown, and B.\ J.\ Enquist,
\textit{Nature} \textbf{400}, 664-667 (1999).

\bibitem {Banavar3}J.\ R.\ Banavar, F.\ Colaiori, A.\ Flammini, A.\ Maritan,
and A.\ Rinaldo, \textit{Phys. Rev. Lett.} \textbf{84}, 4745-4748 (2000).

\bibitem {Banavar4}V.\ Colizza, J.\ R.\ Banavar, A.\ Maritan, and A.\ Rinaldo,
\textit{Phys. Rev. Lett.} \textbf{92}, 198701 (2004).

\bibitem {Banavar}J.\ R.\ Banavar, A.\ Maritan, and A.\ Rinaldo,
\textit{Nature} \textbf{399}, 130-132 (1999).

\bibitem {Kozlowski}J. Kozlovski and M.\ Konarzewski, \textit{Functional
Ecology} \textbf{18}, 283-289 (2004).

\bibitem {Whitfield}J. Whitfield, \textit{Nature} \textbf{413}, 342-344 (2001).

\bibitem {Banavar2}J.\ R.\ Banavar, J. Damuth, A.\ Maritan, and A.\ Rinaldo,
\textit{Nature} \textbf{420}, 626 (2002).

\bibitem {Murray}C. D. Murray, \textit{Proc. Natl. Acad. Sci. USA}
\textbf{12}, 207-214 (1926).

\bibitem {Zamir}M. Zamir, \textit{J. theor. Biol.} 62, 227-251 (1976)

\bibitem {Zamir2}M. Zamir, \textit{J. Gen. Physiol.} \textbf{72}, 837-845 (1978).

\bibitem {Zhi}W. Zhi, Z. Ming, and Y. Qi-Xing, \textit{J. Theor. Biol.}
\textbf{209}, 383-394 (2001).

\bibitem {note1}Note that a fixed total length is not a pertinent constraint,
unless channel cross-sectional areas are bounded; otherwise one could built
channels with infinitely large cross-sections and the minimal resistance value
would be trivialy zero.

\bibitem {Cohn}R. M. Cohn, \textit{Proc. Am. Math. Soc.} \textbf{1}, 316 (1950).

\bibitem {note2}It is also obvious that there are no global maximum, since one
can built a network with infinitely long and indefinitely thin channels (so
that value of $C_{n}$\ is preserved) and whose resistance is infinitely large.

\bibitem {Sherman}T. F. Sherman, \textit{J. Gen. Physiol.} \textbf{78},
431-453 (1981).

\bibitem {LaBarbera}M.\ LaBarbera, \textit{Science} \textbf{249}, 992-1000 (1990).

\bibitem {Roth}A. Roth-Nebelsick, D. Uhl, V. Mosbrugger, and H. Kerp,
\textit{Annals of Botany} \textbf{87}, 553-566 (2001).

\bibitem {Canny}M.\ J.\ Canny, \textit{Phil. Trans. R. Soc. Lond. B}
\textbf{341}, 87-100 (1993).

\bibitem {McCulloh}K. A. McCulloh, J.\ S.\ Sperry, and F.\ R.\ Adler,
\textit{Nature} \textbf{421}, 939-942, (2003).

\bibitem {Kruszewski}P. Kruszewski and S. Whitesides, \textit{J. Theor. Biol.}
\textbf{191}, 221-236 (1998).

\bibitem {Kizilova}N. Kizilova, ICCSA 2004, LNCS 3004, 476-485 (2004).

\bibitem {Xylems}A. H. de Boer and V. Volkov, \textit{Plant, Cell and
Environment} \textbf{26}, 87-101 (2003).

\bibitem {Holbrook}M.\ A.\ Zwieniecki, P.\ J.\ Melcher, C.\ K.\ Boyce,
L.\ Sack, and N.\ M.\ Holbrook, \textit{Plant, Cell and Environment}
\textbf{25}, 1445-1450 (2002).

\bibitem {Kull}U. Kull and A. Herbig, \textit{Naturwissenschaften}
\textbf{82}, 441-451 (1995).

\bibitem {Bohn}S. Bohn, B. Andreotti, S. Douady, J. Munzinger,\ and Y. Couder,
\textit{Phys. Rev. E} \textbf{65}, 061914 (2002).
\end{thebibliography}
\end{document}